\documentclass[useAMS,usenatbib]{mn2e}

\usepackage{graphicx}
\usepackage{txfonts}
\title[The environment of formation as a second parameter]{The environment of formation as a second parameter for globular cluster classification}

   \author[D. Fraix-Burnet, E. Davoust \& C. Charbonnel]{D. Fraix-Burnet$^{1}$\thanks{E-mail: fraix@obs.ujf-grenoble.fr; davoust@ast.obs-mip.fr; Corinne.Charbonnel@unige.ch}, E. Davoust$^{2}$\footnotemark[1] and C. Charbonnel$^{2,3}$\footnotemark[1]\\
$^{1}$Universit\'e Joseph Fourier, CNRS, Laboratoire d'Astrophysique de Grenoble, BP 53, F-38041 Grenoble cedex 9, France\\
$^{2}$Universit\'e de Toulouse, CNRS, Laboratoire d'Astrophysique de Toulouse-Tarbes, 14 av. E. Belin, F-31400 Toulouse, France\\
$^{3}$Observatoire Astronomique de l'Universit\'e de Gen\`eve, 51, chemin des Maillettes, 1290 Versoix, Switzerland}

\begin{document}

   \date{Accepted 2009 June 11.  Received 2009 June 10; in original form 2009 April 29}

\pagerange{\pageref{firstpage}--\pageref{lastpage}} \pubyear{2009}

\maketitle

\label{firstpage}

\begin{abstract}
We perform an evolutionary multivariate analysis of a sample of 54 Galactic globular clusters
with high-quality colour-magnitude diagrams and well-determined ages. 
The four parameters adopted for the analysis are: metallicity, age, maximum 
temperature on the horizontal branch, and absolute V magnitude. 
Our cladistic analysis breaks the sample into three novel groups. An {\it a posteriori} 
kinematical analysis puts groups 1 and 2 in the halo, and group 3 in the thick disc.
The halo and disc clusters separately follow a luminosity-metallicity 
relation of much weaker slope than galaxies. 
This property is used to propose a new criterion for distinguishing
halo and disc clusters.  
A comparison of the distinct properties of the two halo groups 
with those of Galactic halo field stars indicates that the
clusters of group 1 originated in the inner halo, while those of group 2 formed
in the outer halo of the Galaxy.  
The inner halo clusters were presumably initially 
the most massive one, which allowed the formation of more strongly helium-enriched second 
generation stars, thus explaining the presence of Cepheids and of very hot 
horizontal-branch stars exclusively in this group. 
We thus conclude that the ``second parameter" is linked to the environment in which globular clusters
form, the inner halo favouring the formation of the most massive clusters which subsequently
become more strongly self-enriched than their counterparts of the galactic outer halo and disc.
\end{abstract}


%

\section{Introduction}
\label{intro}

Globular clusters are touchstones of astrophysics.galaxies. 
The oldest of them witnessed the formation and early evolution 
of their host galaxies and of their substructures, and their study has historically coloured the different 
scenarios of galaxy formation. However,
as a collective population in a galaxy, they present unsolved problems.
In particular their origin is not firmly established despite the large amount of work devoted to the 
analysis of correlations among their observable properties. It has long been realised that part of the difficulty 
arises from sheer dynamical evolution undergone by these objects since the time of their formation. Indeed, any star
cluster 
is the subject of a long list of erosive mechanisms that operate at different rates 
depending on the cluster's location and orbit within the Galaxy, and on its initial mass
(\cite{DM94}, \cite{GO97}).

Related to these difficulties, the search for a ``second parameter", beyond
metallicity, to explain the distribution of stars along the horizontal branch, 
has met with a limited success. 
\cite{Z93}
suggested that the halo globular clusters
break down into two groups according to their horizontal-branch properties:
the two groups have different ages, kinematics and radial distributions. 
\cite{RYL01} found that an age difference can explain different
horizontal-branch morphologies at a given metallicity.
But \cite{D08}, following the seminal work by \cite{R73} on the impact of 
stellar mass loss on the horizontal-branch morphology, has shown that
considering $\alpha$-element enhancement and metallicity-dependent mass loss 
along the red-giant phase produces similar effects.
On the other hand \cite{LGC07} have found that globular clusters with extended horizontal
branch are more massive than normal clusters and are dominated by random
motions with no correlation between kinematics and metallicity.
Multivariate analyses have been performed by \cite{FP93} who find that
more concentrated clusters have bluer and longer horizontal-branch,
and by \cite{RB06} who find that more massive clusters have a horizontal branch
that extends to higher temperatures.

We can point to several reasons why previous studies have not been
completely successful.  For example, empirically separating 
the clusters into disc and halo populations solely on the basis of one
parameter (metallicity) cannot be satisfactory, as the environment at birth
must influence other cluster parameters. This
was already recognized by \cite{Z85}, who combined metallicity and kinematics 
to establish two major groups (halo versus bulge/disk). The properties of the horizontal
branch have been used, but its usual characterisation by the parameter 
HBR = (B-R)/(B+V+R) is also unsatisfactory. In this
respect, the parameter $T_e$ (maximum effective temperature on the horizontal 
branch) introduced by \cite{RB06} is a welcome innnovation as shown in the present paper.

The classical paradigm describing globular clusters as fairly simple systems of coeval stars of 
homogeneous chemical composition has been seriously challenged recently, and this may bring a crucial piece to the puzzle.
One fundamental characteristic of these systems is their metallicity [Fe/H] (identified as the so-called 
``first-parameter" (\cite{vandenbergh})) that is generally inferred from 
their integrated photometric colours, and that varies strongly from cluster to cluster; in our Galaxy,
globular clusters have [Fe/H] ranging between $\sim$ -2.2 and 0 (\cite{H03}). 
Spectroscopy reveals that, within individual clusters, stars present very homogeneous contents in Fe, but also in $\alpha-$ and s-elements, 
indicating that protoclusters formed from gas pre-enriched in heavy metals (\cite{James04, PC06}). This is in
agreement with the predictions of quantitative models that rule out stochastic self-enrichment in most globular clusters
as a significant contributor to their heavy metals, leaving pre-enrichment as the dominant contributor to metallicity
[Fe/H] (\cite{BH09}).

However globular cluster stars exhibit extremely scattered light-element (Li, F, C, N, O, Na, Mg, and Al) contents 
that are not seen among their field counterparts (see e.g.
\cite{Ca06}). 
This points to early internal chemical evolution (i.e., self-enrichment) in the globular cluster 
driven by first-generation massive and fast evolving stars which polluted with their hydrogen-burning products (among which helium in very 
important quantities) the intracluster gas out of which second generation stars formed (\cite{DCM07}, \cite{PC06}).
Recent findings of double or even multiple stellar populations in the colour-magnitude diagrams of several 
globular clusters, as well as the complexity of the horizontal branch morphology (namely the wide colour 
distribution, i.e., effective temperature, of the stars presently burning helium in their core) constitute 
further evidence for internal evolution (\cite{P09}). All these features can indeed be related to the presence of a second generation 
of He-enriched stars. Importantly, the star formation history depicted by these features seems to vary from cluster to
cluster (\cite{Milone2008}) 
in a way which is still far from being understood.
However we have now firm evidence that Galactic globular clusters have undergone internal 
chemical evolution and complex star formation histories during their infancy that shaped their properties 
and in particular their present total mass (\cite{DCM07}). 
This new paradigm has opened a novel route for a better understanding of the origin and history 
of globular clusters.

It thus appeared to us that a multivariate analysis which simultaneously 
takes into account any cosmic variance due to evolving physical conditions and 
groups objects according to environment of formation would 
be very valuable. Cladistics provides such a methodology. 
It differs from other clustering analyses in that it focuses on evolution 
within and between groups rather than on similarities between objects (\cite{WSBF91}). 
Cladistics is very commonly used in evolutionary biology and has been 
pioneered in astrophysics by \cite{jc1,jc2,DFB09}  and successfully 
applied to the dwarf galaxies of the local group (\cite{FCD}). 

This paper presents a multivariate analysis based on the method of cladistics of a large sample of Galactic
globular clusters.
After presenting the data and the method of analysis (Sect.~\ref{data}),
we describe the three groups found by the cladistic analysis
(Sect.~\ref{3groups}), and discuss two important results,
evidence for self-enrichment (Sect.~\ref{enrichment}) and
a possible luminosity-metallicity relation (Sect.~\ref{mv-fe}).
We then compare the properties of the three groups with those
of Galactic halo field stars (Sect.~\ref{comp}), before proposing
a scenario for the formation of the three groups (Sect.~\ref{origin}).

\section{Data and method of analysis}
\label{data}

The choice of parameters is a crucial step in any multivariate analysis.
\cite{DM94} have shown that the manifold of Galactic globular cluster
properties has a dimension larger than 4, but that a subset of parameters
linked to morphology and dynamics forms a three-dimensional family. 
Including properties of the stellar populations (e.g. a horizontal-branch
parameter or metallicity) will increase the manifold by 1 or 2 dimensions
(\cite{FP93}). Finally, using a large number of photometric and structural parameters, 
\cite{RB06} found that 4 eigenvectors account for 79\% of the total sample variance.

Taking advantage of this indication, we selected the following four parameters for analysis:
relative ages, 
metallicity ([Fe/H]), absolute V magnitude ($M_v$), and maximum effective temperature ($T_e$) on the horizontal branch.
The age parameter is related to the secular evolution of the stellar populations.
[Fe/H] reflects the chemical composition of the environment when and where globular clusters 
formed and is the ``first parameter" for the horizontal branch morphology
$M_v$ is a structural parameter that measures the present total baryonic mass of the globular clusters\footnote{The
absolute magnitude 
$M_v$ of globular clusters in the Milky Way spans a vast range ($-1.7 \leq M_v \leq -10.2$, \cite{H03}), and
reflects a 
large mass range ($10^3 - 10^6 M_{\odot}$, \cite{Mv05}).}.
Finally $T_e$ is a measure of both the pristine chemical composition of the protocluster 
([Fe/H] being the first parameter) and of the helium enrichment during early internal chemical evolution, 
since stars with higher helium content are expected to reach higher effective temperatures on the 
horizontal branch (\cite{Da02, RB06}).
The last three quantities describe truly intrinsic properties of globular clusters. 
As the age parameter evolves in all clusters, it cannot be used to classify them at 
the same level as the other three parameters. We thus gave it a lower weight in the 
cladistic analysis (see Appendix A).

We performed our analysis using the large sample (54 objects) of \cite{RB06} based on homogeneous Hubble Space
Telescope 
photometry. We used the $T_e$ values obtained uniformly from this database by \cite{RB06}, as well as the relative ages and $M_v$ 
values they adopted (i.e., taken respectively from \cite{DE05} and the 2003 on-line revision of  \cite{H03}).
We did not include more parameters in the cladistic analysis 
as we preferred to avoid the unwanted effect of redundancies, which give more weight to correlated parameters. 
Note also that we did not use any kinematical information in the cladistic analysis.
However we used other parameters {\it a posteriori} to characterize the different
groups found by the cladistic analysis: the radial velocities and structural
parameters were taken from the 2003 on-line revision of \cite{H03}, the orbital parameters from
\cite{D99,D03,D07}. 
In this respect, the distinct orbital properties of group 3 found a posteriori (see \S 3) are independent of the
methodology.

More recent age estimation was published by \cite{MarinFranch} after most of this project was completed, but for
only 35 out of 54 globular clusters of our sample. Using these produces an inhomogenous data set, from different
sources, relying on
different values of Fe/H. The ages of  \cite{MarinFranch} have been determined from colour-magnitude diagrams and values
of
Fe/H which are different from those of \cite{RB06}. \cite{MarinFranch} themselves point out the importance of using a
homogeneous set of Fe/H to derive the ages. Nevertheless, it is instructive to perform analyses using both
sets and compare the results, so as to determine how sensitive they are to the specific choice of parameter values. One then has the problem of combining two sets of ages, and it
 is not obvious how this should be done.
 It turns out that 9 out of 11 globular clusters that calibrate relative 
 ages of \cite{MarinFranch} are in common with \cite{DE05}, and using the relative ages of \cite{DE05}
 for these 9 globular clusters gives a mean value of 1.0055 instead of 1.00, which is fine. 
 However, comparing the ages of all the globular clusters in common suggests a non-linear 
 systematic effect, which should perhaps be taken into account.
 Furthermore, we cannot simply convert each set of relative ages
 to absolute ages with the zero-point of each set, because the two
 zero-points are rather different  (11.2Gyr for \cite{DE05} and 12.8Gyr for  \cite{MarinFranch}). We thus simply used
the
relative ages without any attempt at homogenizing them, and the zero-point of one author for all ages. This additional
analysis is compared to the main one in Sect.~\ref{addanal}.

The multivariate analysis was performed using the method of cladistics.
In short, the method works as follows. One first builds a matrix with values of the
four parameters for each clusters. The values must be discretized, and the
number of bins (here 10) depends on the resolution one wants for the analysis.
One then chooses a cluster which represents the most unevolved state,
in the present case the metal-poorest cluster (NGC 6934), and the software
classifies all the other clusters in order of increasing diversification
of properties (in other words, by increasing distance in the manifold
of parameters). Clusters which are diverging from the original cluster
in the same direction are put on the same branch. We refer to Appendix A and \cite{jc1, jc2, FCD, DFB09}
for more details on the principles of the method.

\section{Three groups of clusters}

\subsection{The main tree}
\label{3groups}

\begin{figure}
   \centering
   \includegraphics[width=6.5 true cm]{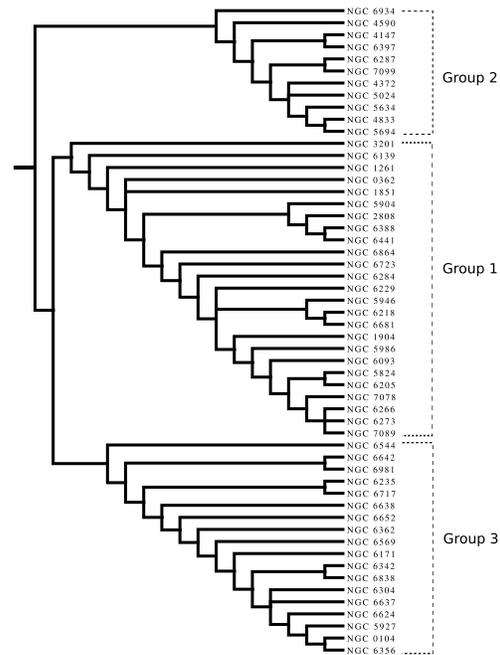}
   \caption{The tree resulting from the cladistic analysis. The sample
break into three distinct branches. Group 1 is composed
of inner halo clusters, group 2 of outer halo clusters, and group 3 of
disc clusters}
              \label{treefig}%
    \end{figure}

The main result of our analysis is presented in the form of a tree structure,
a usual form of representation in graph theory.  The properties of the sample
can be read from the structure of the tree shown on Fig.~\ref{treefig}. 

The tree has been rooted with group 2 which has the lowest metallicity on average, 
and as such is supposedly made of more primitive (or "ancestral") material. 
The tree divides into three main branches, which define
three groups with quite distinct properties.
Age increases roughly monotonically
along each branch, as expected. There are some subbranches
sharing similar values of the adopted parameters, such as NGC 5904, 2808, 6388 and 6441,
and which set them apart in the four-dimensional space of parameters, 
implying that they might in fact belong to a fourth small group with 
properties similar to those of group 1. 
Specialists will immediately remark that the latter three clusters
indeed share peculiar properties, in particular a very Helium-rich stellar
population (\cite{P08}).  We emphasize that the Helium abundance is not
one of the parameters included in the cladistic analysis: it thus must
influence one way or another the four parameters used in the analysis.

\begin{table}
 \centering
 \begin{minipage}{\columnwidth}
  \caption{Properties of the three groups of globular clusters. No age estimate is available for NGC 6139, 6229,
6304, 6388, 6441, 6569 and 6642. The middle part of the Table gives 
the correlations:  + means a correlation, - means an anticorrelation, x no correlation. 
The orbital properties presented in the lower part of the Table are only available for a subset of each group. Numbers in brackets are rms dispersions}
     \label{tableprop}
  \begin{tabular}{llll} 
  \hline
          & Group 1 & Group 2 & Group 3\\ 
          & inner halo&outer halo&thick disc\\
 \hline
Number of clusters& 25 & 11 & 18\\
$R_{gc}$ (kpc)& 9.4 (7.4) & 12.9 (8.0) & 4.2 (2.9)\\ 
$Z$ (kpc) & 4.8 (4.5) & 8.6 (7.6) & 1.9 (2.0) \\
Fe/H & -1.40 (0.35) & -1.92 (0.16) & -0.92 (0.35)\\
$M_v$  (mag)  & -8.5 (0.7) & -7.6 (0.6) & -7.1 (0.9)\\
$V_{rot}$ (km/s) & -7.     &   +46.   & +119.\\ 
$\sigma$ (km/s) & 120 (107) & 151 (107) & 69 (74) \\ 
Age & 9.98 (0.96)$^a$ & 11.17 (0.70) & 10.18 (0.48)$^b$  \\
 \hline
Age- $log(T_e)$   & + & x & -\\
Age- [Fe/H]     & - & - & +\\
Age- $M_v$        & x & x & x\\
$log(T_e)$ - [Fe/H] & + & x & +\\
$M_v$ - [Fe/H]  & + & + & +\\
$M_v$ - $log(T_e)$ & + & x & x\\
 \hline
Number of clusters& 12 & 8 & 5\\
$P$ (Myr)    & 353 (212) & 391 (254) & 142 (35) \\ 
$E$ ($10^2km^2/s^2$) & -691 (341) & -649 (339) & -1027 (111)\\
$e$          & 0.63 (0.18) & 0.54 (0.18)& 0.21 (0.10)\\ 
$R_a$ (kpc)& 17 (12)& 19 (13)& 6 (1.3)\\ 
$Z_{max}$  (kpc) & 7.2 (6.2)& 9.6 (8.5) & 1.5 (1.0)\\ 
$\Psi$ (deg) & 32 (12) & 38 (17) & 21 (15)\\
$L$          & 886 (802) &  941 (575) & 866 (314)\\ 
$|W|$   (km/s) & 100 (78)$^c$ & 77 (40) & 34 (16)$^d$\\ 
$\Pi$ (km/s) & +19 (141) & -33 (124) & -3 (22)\\
$\Theta$ (km/s) & +15 (121) & +89 (129) & +170 (33)\\
\hline
\end{tabular}
\end{minipage}
$^a$ average of 21 values; 
$^b$ average of 15 values;\\ 
$^c$ average of 13 values;
$^d$ average of 6 values.
\end{table}


The properties of the three groups are presented in Table~\ref{tableprop}.  
The first 7 rows give the characteristics of the 3 groups:
number of clusters in each group, mean distance from the Galactic center
$R_{gc}$, height above the Galactic plane $Z$, metallicity [Fe/H],
absolute magnitude ($M_v$),
mean velocity of rotation $V_{rot}$ in the Galactic plane (computed
with the equations given by \cite{FW80} using a velocity of the Sun of 220 km/s), 
radial velocity dispersion $\sigma$. 
The next 6 rows describe the correlations between the four parameters. 
The bottom part of the Table lists orbital parameters taken from \cite{D99,D03,D07}, which
were not available for all clusters of each group: successively number of clusters, 
period of rotation $P$, total energy $E$, 
eccentricity of the orbit $e$, apocentric distance $R_a$,
maximum distance reached above the Galactic plane $Z_{max}$, 
inclination angle with respect to the Galactic plane $\Psi$,
angular momentum $L$, and finally the velocity components
in cylindrical coordinates: vertical velocity $|W|$,
radial velocity $\Pi$ and tangential velocity $\Theta$.  

Paired t-tests showed that the differences of the means of the groups taken two by two is not equal to 0 ($p < 0.05$)
for the parameters $log(T_e)$, $R_{gc}$, [Fe/H], $e$. This is also the case between group 1 and group 2 and between
group 1 and group 3 for $M_v$ and $\Theta$, and between group 2 and group 3 for $Z$ and age. There is also evidence
($0.05 < p < 0.1$) for different means of  the latter two groups for $P$, $E$, $R_a$, $Z_{max}$ and $|W|$, as well as
between group 1 and group 2 for age and between group 1 and group 3 for $|W|$. We emphasize that the rotational
properties of Galactic globular clusters are very uncertain, since they are derived from projected radial velocities or
numerical simulations of orbits in a model Galaxy.

These properties show that the first two groups belong
to the halo population of clusters, while the third group belongs to the thick disc population.
Hereafter, the thick-disc clusters will simply be called disc clusters.
The average velocity of rotation of group 1 and 2 together is $V_{rot}$ = 9 km/s. 
Group 3 is confined to the Galactic plane, and has a high $V_{rot}$ and low $\sigma$.
If we separate group 3 into two subgroups of equal size according to their distance
from the Galactic center, we find that $V_{rot}$ is 88 km/s for the inner 
subgroup ($R_{gc} < 3$ kpc) and 187 km/s for the outer subgroup.
There is also evidence that group 3 has a shorter $P$, lower $e$, $\Psi$ and $Z_{max}$, 
and no radial motion,
as expected from clusters that partake in the overall rotation of the disc. 

One cluster has certainly been misclassified. NGC 6981 is in group 3
although it is at $Z$ = 9.1 kpc and has a low velocity of rotation.
Since NGC 6981 is a borderline cluster in all the Figures, the value of one of the four
parameters may be erroneous.  Indeed, raising $M_v$ from --7.04
to --7.27 brings it into the next bin in our cladistic analysis (which requires
that the data be discretized into a limited number of bins), and running
the cladistic analysis again moves the cluster to group 2.  
Moving any of the other three parameters by one bin and redoing the
cladistic analysis does not change the status of the cluster. 
Since we found no reason for an erroneous $M_v$, we left it in group 3.

Another possible discrepant cluster is NGC 6266, which is in group 1, but
which, according to \cite{D03}, belongs to a rotationally supported system, 
on the basis of its kinematics (but without precise orbital determination), 
despite its low metallicity. However, it is on the same
subbranch as (and undistinguishable from) NGC 7089, which definitely belongs
to the halo, according to our analysis and that of \cite{D99}. In addition, as pointed out by the referee, an
isotropic distribution of orbits will statistically produce one or several ones in or near the Galactic plane.
We are thus confident that NGC 6266 belongs to the halo.

We now compare the statistical properties of the two halo groups with those of the
Galactic halo field stars. 
The dichotomy of the Galactic halo stellar population has been suspected for some time. 
The most quantitative study in that respect, that of \cite{Car07}, clearly identifies two 
broadly overlapping structural components corresponding to an inner and an outer halo.
Stars of the inner halo are in highly eccentric orbits, in slightly prograde rotation,
and have an average metallicity of [Fe/H] = --1.6. The outer halo stars have a 
uniform distribution of eccentricities, are in highly retrograde orbits, and of
lower metallicity [Fe/H] = --2.2.  These properties are also among those that distinguish
the two groups of halo clusters, and indicate that the clusters
of group 1 may have originated in the inner halo: they have higher eccentricities
and metallicities than group 2, while group 2 formed
in the outer halo: they have the largest $R_{gc}$, $R_a$, $Z$ and $Z_{max}$, 
lowest metallicities; $\Theta$ is positive, but not significantly so.

\subsection{Additional analysis using an inhomogeneous set of ages}
\label{addanal}

We also applied our analysis to the sample of 54 globular clusters with composite
ages as explained in Sect.~\ref{data}. It leads to roughly the same three groups
as before, with several obvious halo clusters moving into G3 (which
is in principle composed of thick-disk globular clusters), and two globular clusters of G1 moving
into G2, among them NGC 6218 which has a Cepheid (and should thus be in G1).  
In other words, the number of misclassified globular clusters rises from one (NGC 6981) 
to only about 6. This indicates that our groups are fairly robust. Of course, their detailed contours depend 
on the choice of data and we believe that the slight discrepancy is due to the inhomogeneous
data of this additional analysis: 
the ages of \cite{MarinFranch} have been determined from colour-magnitude
diagrams and values of Fe/H which are different from those of
\cite{RB06}, while logTe comes from \cite{RB06} and is derived from diagrams using the corresponding set
of Fe/H values.

\section{Evidence in favour of self-enrichment}
\label{enrichment}

Self-enrichment by a first generation of stars is frequently advocated
to explain the chemical anomalies in globular clusters (e.g. \cite{PC06}) in relation with horizontal-branch
morphology (\cite{DaC08}).
We list below distinctive properties of the three groups of globular clusters 
pointing to such a process.

\begin{figure}
   \centering
   \includegraphics[width=6.5 true cm]{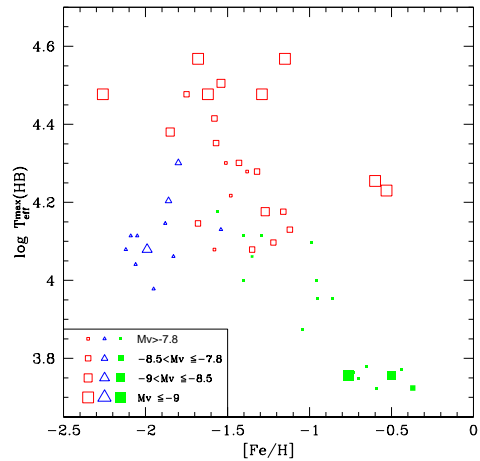}
   \caption{Metallicity-logTe diagram with symbol size indicating visual magnitude. The halo clusters are represented by open symbols: squares for group 1 and triangles for group 2. The disc clusters (group 3) are represented by full squares
} 
    \label{fe_tfig}%
    \end{figure}

The well-known correlation between metallicity and extent of the 
horizontal branch is clearly present in Fig.~\ref{fe_tfig}: the more metallic
the cluster, the less extended its horizontal branch at a given metallicity. 
Also, the more luminous (that is the more massive) clusters tend to 
have more extended horizontal branch (\cite{RB06, LGC07}).  This latter point is 
easily understood in the self-enrichment framework, in the sense that more 
massive globular clusters retain the helium-rich ejecta of massive polluter 
stars in their deeper potential well more efficiently than less massive 
globular clusters . 

The crucial new information brought by our analysis in this context is that 
the present mass of the clusters seems to depend on their origin, the inner 
halo clusters being presently more massive than their outer halo counterparts 
(Fig.~\ref{fe_tfig}). In fact, in order to have extreme light-element abundance 
patterns, and in particular extreme O-Na anticorrelation 
(\cite{Ca06}), which is linked to the extent of the horizontal branch
(\cite{Ca07}), the inner halo clusters must 
have been even more massive in the past, before they lost a huge number (96\%) 
of first-generation low-mass  stars in their early dynamical evolution (\cite{DCM07}). 
We thus expect the mass difference between the inner and outer halo globular 
clusters to have been even larger in the past.

\begin{figure}
   \centering
   \includegraphics[width=6.5 true cm]{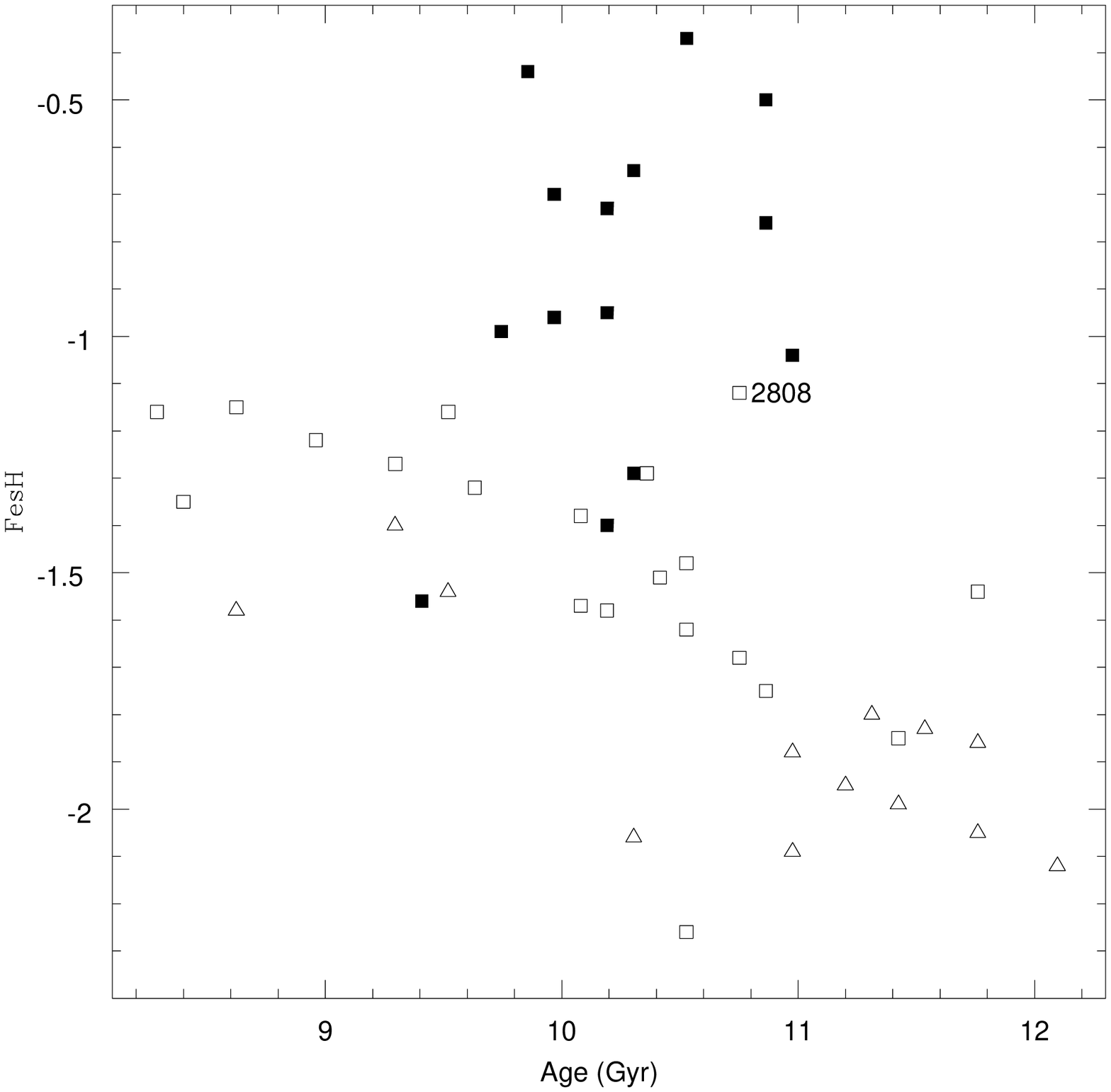}
   \caption{Age-metallicity diagram. Same symbols as Fig.~\ref{fe_tfig}.
No age estimate is available for NGC 6139, 6229, 6304, 6388, 6441, 6569 and 6642. Metallicity decreases with age in
the halo clusters 
while it very marginally increases with age in the disc clusters} 
              \label{age_fe2fig}%
    \end{figure}

The disc and halo clusters are well separated in the age-metallicity diagram,
shown in Fig.~\ref{age_fe2fig}.  For the halo clusters, metallicity decreases with age.
For the disc clusters, on the contrary, it marginally increases with age, if at all.  
The spread in age is 4 Gyr for the halo component and only 1.5 Gyr for the disc one.
The figure confirms that the metallicity of NGC 2808 should indeed be about 0.5dex
lower. The two other He-rich clusters of group 1 do not appear on this plot because no
age estimate is available for them, but their metallicity is indeed about 0.5dex
higher than the highest metallicity of the rest of the halo population.  
This supports the claim by 
\cite{CDa08} and \cite{PC06} (see also \cite{DCM07}) that He-enrichment must be
associated with the build-up of abundance anomalies of light elements during the phase of self-enrichment.

The presence of multiple stellar populations in the colour-magnitude diagram
is another evidence for self-enrichment. Unfortunately, only three of the
clusters of the sample, NGC 1851, 2808 and 6388, all belonging to group 1,
are known to have such a feature (\cite{P09}). We predict that two other known
such clusters, NGC 5139, 6656 which are both metal-poor and massive, 
are also inner halo globular clusters.

Additional clues to the self-enrichment scenario can be gathered from the 
RR Lyrae and Cepheid contents of the three groups. Globular clusters have 
historically been divided into two groups (Oosterhof I and II) according 
to the properties of their RR Lyrae stars. Using the compilation of \cite{CC01}, 
we find that the two Oosterhof types are equally present in group 1 and 2.  
More interestingly, we find that the distribution of periods of RR Lyrae stars 
in group 2 is more sharply peaked than the corresponding distribution for 
group 1, which presents a minor secondary peak at a higher period.  
The narrower period distribution in group 2 implies a small dispersion in 
mass loss along the red-giant branch (\cite{CDa08}), while the wider distribution in 
group 1 implies several generations of stars, each with a narrow distribution 
of mass loss along the red-giant branch, and with increasing helium abundance (\cite{DaC08}), 
reinforcing the necessity of self-enrichment.  

Furthermore all the population II Cepheids are found in clusters of group 1, 
and none in clusters of group 2.  Population II Cepheids result from the 
evolution of post-horizontal branch stars which start from the higher 
temperatures of the zero-age horizontal branch and move toward the asymptotic 
giant branch or leave that branch on rapid blueward loops (\cite{W02}); 
this explains their absence in halo clusters with low $Te$.

We thus reach the conclusion that the inner halo favours the formation of
very massive clusters, which retain more easily the products of first-generation
stars and thus become more strongly self-enriched, giving rise to more extended horizontal branches.

\section{A luminosity-metallicity relation for globular clusters?}
\label{mv-fe}

\begin{figure}
   \centering
   \includegraphics[width=6.5 true cm]{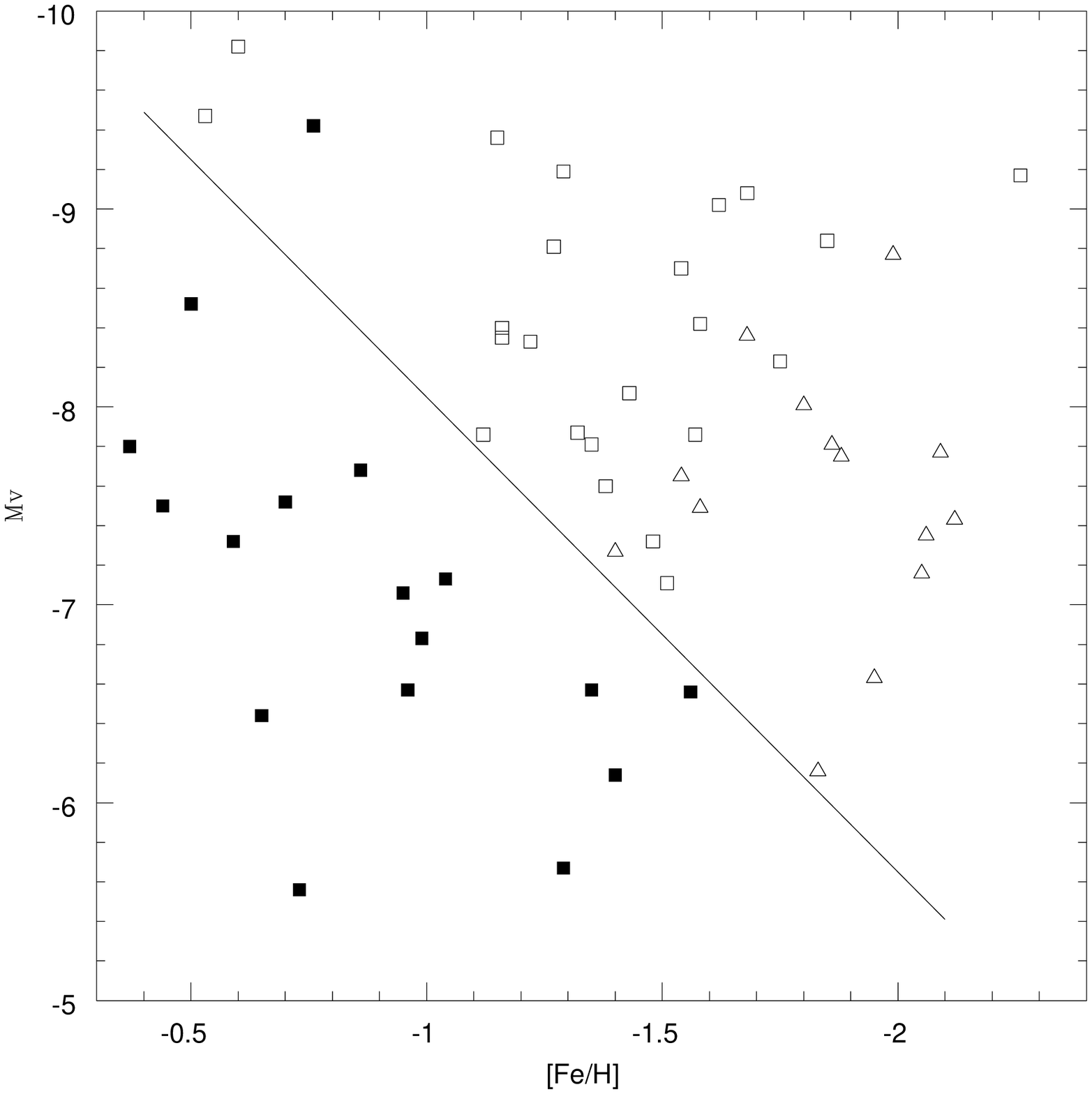}
   \caption{Metallicity-$M_v$ diagram. Same symbols as Fig.~\ref{fe_tfig}. 
The solid line is the limit between disc and halo clusters according
to our criterion ($M_v$ = --2.4$\times$Fe/H -- 10.45). The disc cluster above
this line is NGC 104} 
    \label{fe_mvfig}%
    \end{figure}

A mass-metallicity or luminosity-metallicity relation is found among galaxies, 
but is not expected and has not been found in the Galactic globular cluster 
system (\cite{DM94}), nor in numerical simulations of globular cluster 
formation (\cite{KG05}). However, it has been found, in the form of a 
"blue tilt", in colour-magnitude diagrams of globular cluster systems in bright 
ellipticals (\cite{BS06,H06,H09,Mieske}). Such a relation, where metallicity increases with Mv, 
is present in our sample (Fig.~\ref{fe_mvfig}), if we consider the disc and halo clusters 
separately. 

We have plotted a line separating the disc and halo clusters in 
Fig.~\ref{fe_mvfig}, which can be used as a criterion for distinguishing the two types of 
clusters, in conjunction with other criteria, since the separation is not 
perfect (NGC 104 is a notable exception). The disc clusters are on average 
fainter than the halo ones by about 3 magnitudes at a given metallicity. 
Correlation lines were adjusted to the two subsamples: they have comparable 
slopes of -2.0 and -2.8 for the halo and disc clusters respectively, 
with admittedly low correlation coefficients of 0.53 and 0.32. 
These slopes are much lower than that of -6.75 predicted and found for 
the dwarf galaxies of the Local Group (\cite{DS86}). They were however probably 
much steeper when the globular clusters formed, since we expect some of these 
objects to have lost a large fraction of their initial mass in the 
self-enrichment framework.

A remarkable property of our disc clusters is that they extend to metallicities
lower than the conventional limit of [Fe/H] = -- 0.8 (\cite{Z85}). 
This is not very surprising {\it per se}, since \cite{D99} have found three 
metal-poor Galactic globular clusters with thick-disc 
kinematics, and in M31 there is also evidence for metal-poor globular clusters 
with disc kinematics (\cite{M04}, although see \cite{FP05}).

If we divide group 3 along the conventional limit, we find that the metal-rich 
clusters have 
$V_{rot}$ = 184 km/s whereas the other ones have $V_{rot}$ = 71 km/s,
rather low, but still significantly larger than that of the halo clusters.
We have checked that this low mean velocity is not due to one cluster in particular.
The two subgroups do not distinguish themselves otherwise; in particular they 
have the same spatial distribution (same mean $R_{gc}$ and $Z$). 
Two of the low-metallicity disc clusters, 
NGC 6171 and 6362, have orbits determined by \cite{D99}, which confirm 
that they do belong to the disc population. In fact \cite{D99} state that their most 
significant result is to have shown the existence of metal-poor clusters
with orbits consistent with the thick-disc motion. Our analysis confirms this finding.

The second important result of the present paper is that the disc and halo globular
clusters should not be separated on the basis of metallicity, but rather
of a multivariate analysis, taking into account other parameters.
We propose to use the magnitude at a given metallicity as a rough criterion,
with a limit such that $M_v = -2.4\times [Fe/H] - 10.45$, together
with other criteria, such as location in the Galaxy and
velocity of rotation.  

\section{Comparison with other studies and with halo field stars}
\label{comp}

Before interpreting the differences between the three groups in terms of 
formation history, we compare them to the traditional disc, young halo and old 
halo groups of \cite{Z93}, to emphasize that they are rather different. 
There are 4, 4 and 1 young halo clusters in group 1, 2 and 3 respectively, and 
the numbers are 17, 10 and 8 for the old halo clusters. Two of Zinn's disc 
clusters (NGC 6388 and 6441) are in our group 1. Since they are located 
near the Galactic centre and have no determined orbits, it is not possible to 
decide if they kinematically belong to the disc or the halo. Furthermore, as
mentioned in Sect.~\ref{3groups}, these two clusters might in fact belong to 
a fourth small group, maybe of Galactic bulge clusters.  Turning to the 
metallicity vs HBR diagram, we find that group 1 extends 
to lower metallicities at a given HBR than the old halo clusters, and that 
there are clusters of group 1 and 2 among the old halo clusters with the 
reddest HBR. Comparing our groups to those of \cite{LGC07}, we find that 
all the clusters of their group with extended horizontal branch are in our 
group 1, except NGC 4833 (which is in our group 2).

If we now compare (Table~\ref{tableharris}) our grouping with that
of \cite{H01} (see his Table 1.6), we find that G3 dominates in metal-rich clusters class (MRC)
and G1 and G2 dominate in metal-poor cluster class (MPC). In MPC alone there is no clear separation
in R$_{gc}$ between G1 and G2, while G3 tends to be in the inner regions. G2 tends to be among the more metal-poor
globular clusters and
tends to be at larger galactocentric distances. In summary, the MPC/MRC dichotomy corresponds
roughly to our halo/disk separation, and our G2 populates the very metal-poor and distant MPC. Table~\ref{tableharris}
also shows that a classification based on arbitrary criteria
does not quite retrieve the groups obtained with a multivariate
analysis.

\begin{figure}
   \centering
   \includegraphics[width=6.5 true cm]{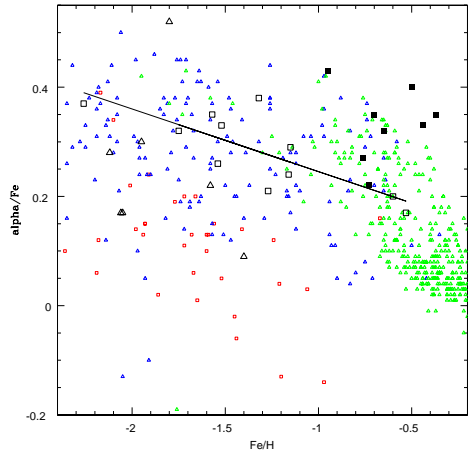}
   \caption{Metallicity versus light-element abundance. Same symbols as previous figures. The line is a least-squares
fit to the globular clusters of group 1. Small triangles are disc (green) and halo (blue) field stars. Small red
open squares are dwarf galaxies of the Local Group (from Venn et al. 2004)
} 
    \label{fig5}%
    \end{figure}

We now compare the content in
$\alpha$-elements of the two groups of halo globular 
clusters with those of halo field stars. Several studies have shown that the 
field stars in the outer halo, identified through their kinematical or orbital 
parameters, tend to have lower and/or more dispersed relative abundances in 
$\alpha$-elements (\cite{SB02,G03}), which points to a difference in star 
formation rates of their birth environment: the $\alpha$-elements are indeed almost 
exclusively provided by core-collapse supernovae, which arise from ephemeral 
massive stars, while Fe is essentially produced in type Ia supernovae, which 
arise from stars with longer lifetimes. We find marginal evidence for a 
similar difference between groups 1 and 2 (see Fig.~\ref{fig5}). [$\alpha$/Fe] decreases 
with increasing [Fe/H] in group 1, while it is more dispersed and shows no 
clear trend with [Fe/H] in group 2. This confirms a similar 
origin for the $\alpha$-elements in globular 
clusters and field stars of the halo,
which thus has to occur prior to protocluster formation. 

\begin{table}
 \centering
 \begin{minipage}{\columnwidth}
  \caption{Comparison between Table 1.6 of Harris (2001) and our grouping by number of clusters for each class.}
     \label{tableharris}
\begin{tabular}{llrrrr}
\hline
           &  & Total & G1  & G2 & G3 \\
\hline
MRC   &   All [Fe/H] $>$ -1                & 14 &  2 & 0 &12 \\
MRC   &   R$_{gc}$ = 0 - 4 kpc               &  9 &   2& 0 &  7 \\
MRC   &   R$_{gc}$ = 4 - 9 kpc               &  5 &   0& 0 &  5\\
           &                                           &     &  &  & \\
MPC   &   All [Fe/H] $<$ -1                & 40 & 23 &11  &  6 \\
MPC   &   R$_{gc}$ = 0 - 4 kpc               &  10&   6&  1 &   3\\
MPC   &   R$_{gc}$ = 4 - 8 kpc               &  11&   5&  4 &   2\\
MPC   &   R$_{gc}$ = 8 - 12 kpc             &7   & 6   & 1 & 0 \\
MPC   &   R$_{gc}$ = 12 - 20 kpc           & 7  &  4  &2 &   1 \\
MPC   &   R$_{gc}$ $>$ 20 kpc                   & 5  & 2  &3   & 0 \\
           &                                            &    &  &  & \\
MPC   &   -2.30 $< $[Fe/H] $\le$-1.85   & 9 &1   &8   & 0\\ 
MPC   &   -1.85 $< $[Fe/H] $\le$-1.65   & 6 &4   &2   & 0   \\ 
MPC   &   -1.65 $< $[Fe/H] $\le$-1.50   & 8 &6   &1   & 1\\ 
MPC   &   -1.50 $< $[Fe/H] $\le$-1.32   & 8 &5   &0   & 3\\ 
MPC   &   -1.32 $<$ [Fe/H] $\le$-1.00   & 9 &7   &0   & 2\\ 
MPC   &   All [Fe/H] $<$ -1.70            &27&  20 & 1&6 \\
          &                                             &  &  &  & \\
MPC   &   HBR$>$0, R$_{gc}$ $>$ 8kpc  & 13 &  7  & 5    & 1   \\                 
MPC   &   HBR$<$0, R$_{gc}$ $>$ 8kpc & 5 &  5  &0    & 0 \\                   
MPC   &   HBR$<$0     &18&  6  & 0 &    12    \\           
\hline   
\end{tabular}
\end{minipage}
\end{table}

\section{Origin of the three groups of globular clusters}
\label{origin}

It is generally assumed that the Galaxy assembled in a hierarchical fashion 
from collapsing haloes of dark matter (\cite{B98}. Small proto-galactic clumps 
formed first, from initial small-scale density fluctuations, and collapsed in 
a dissipationless way (because their gas was quickly consumed or blown away), 
or else they merged and grew in size to form larger clumps which spiralled 
toward the inner regions of the Galaxy by dynamical friction and experienced 
a dissipational collapse. This scenario, or variants of it, first proposed by 
\cite{SZ78} and verified by cosmological simulations(\cite{BC01}), has 
repeatedly been invoked to explain the properties of the Galactic globular 
clusters and of the stellar halo. Furthermore, cosmological simulations have 
shown that globular clusters can form in the densest regions of collapsing 
subhaloes of dark matter and giant molecular clouds, when the clouds reach a 
critical density and are under a high external pressure. The mass distribution 
function of the clusters is similar to that of the clouds (\cite{KG05}).

It has often been proposed that some clusters, in particular
those identified as young halo clusters by \cite{Z93}, which tend to be in the outer halo and 
counter-rotating, were formed by accretion and disruption of satellite 
galaxies. But the chemical homogeneity of the halo, as well as substantial 
differences in chemical composition between field stars in the halo and dwarf 
spheroidal galaxies, argue against the accretion scenario 
(\cite{SB02,PVI05,G07}).

The properties of our three groups of clusters can be interpreted in the 
following way, without resorting to an external origin for any of the groups:

-- The clusters of the outer halo (group 2) formed during the initial 
dissipationless collapse of the proto-galaxy, from material already polluted 
by earlier generations of stars, but not well homogenized and thus 
inhomogeneous in $\alpha$-elements. Contrary to the outer halo stars, they lost 
their initial average retrograde rotation by dynamical friction and 
gravitational encounters. As suggested by the referee, this
group could also have originated in some "pregalactic dwarfs" (i.e. 
metal-poor, gas-rich satellites that soon afterward began hierarchical 
merging).

-- The clusters of the inner halo (group 1) formed later, during the 
dissipational phase of Galactic collapse, which continued in the halo after 
the formation of the thick disc and its globular clusters. Since the formation 
of group 1 occurred later, the molecular clouds from which they formed had time 
to grow by accretion of smaller clumps. These clouds were already enriched at 
the same level in $\alpha$-elements. Thanks to the strong potential well in the 
clusters (as evidenced by their high central velocity dispersions), the He-rich 
ejecta of first generation massive stars were not blown away and found their 
way into a more strongly helium-enriched second generation of stars, 
favouring the production of hot horizontal-branch and Cepheid stars. 

-- As indicated by their short range in age (1.5 Gyr, see Fig.~\ref{age_fe2fig}), the disc 
clusters (group 3) formed in a more rapid fashion than the two other groups, 
before many clusters of group 1. This could presumably be due to the higher 
densities and external pressure in the thick disc. This group shows significant 
average 
prograde rotation, because the dissipational collapse of the disc conserved 
angular momentum. The metal-poor disc clusters seem to rotate more slowly and 
have larger eccentricities and inclinations than the metal-rich disc clusters.  
Since there is no significant age difference between the two subgroups, we 
assume that the metal-poorer clusters formed further away from the Galactic 
plane, and thus retained a larger vertical velocity component.

\section*{Acknowledgements}
Paired t-tests were performed with the VOStat software (http://vo.iucaa.ernet.in/~voi/VOStat.html). We thank Asis
Chattopadhyay for advice. We would like to thank the referee, W. Harris, for his constructive comments
and suggestions.

\appendix

\section[]{Cladistics applied to globular clusters}
\label{cladtech}

Multivariate clustering methods compare objects with a given measure and then gather them according to a proximity
criterion. Distance analyses are based on the overall similarity derived from the values of the parameters describing
the objects. The choice of the most adequate distance measure for the data under study is not unique and remains
difficult to justify a priori. The way objects are subsequently grouped together (this is called the linkage) is also
not uniquely defined. Cladistics uses a specific measure that is based on characters (a trait, a descriptor, an
observable, or a  property, that can be given at least two states characterizing the evolutionary stages of the object
for that character) and compares objects in their evolutionary relationships. Here, the "distance" is an evolutionary
cost. Groupings are then made on the basis of shared or inherited characteristics, and are most conveniently represented
on an evolutionary tree. 

Character-based methods like cladistics are better suited to the study of complex objects in evolution, even though the relative evolutionary costs of the different characters is not easy to assess. Distance-based methods are generally faster and often produce comparable results, but the overall similarity is not always adequate to compare evolving objects. In any case, one has to choose a multivariate method, and the results are generally somewhat different depending on this choice (e.g. \cite{Buchanan}). However, the main goal is to reveal a hidden structure in the data sample, and the relevance of the method is mainly provided by the interpretation and usefulness of the result. In the present paper, the use of cladistics is justified a priori by the evolutionary nature of globular clusters, and a posteriori by the strong astrophysical significance of the grouping found.

In astrophysics, cladistics has already been applied to galaxies because they can be shown to follow a transmission with
modification process when there are transformed through assembling, internal evolution, interaction, merger or stripping
(\cite{jc1,jc2,FCD,DFB09}). For each transformation event, stars, gas and dust are transmitted to the new object with
some modification of their properties. For globular clusters, interactions and mergers are probably rare. It was
previously thought that once they assembled, only the stellar ageing would affect their properties. Nowadays, we have
firm evidence that internal evolution can create another generation of stars, and clusters can lose mass. Basically, the
properties of a globular cluster strongly depend on the environment in which it formed (chemical composition and
dynamics), and also on the internal evolution which includes at least the ageing of its stellar populations. To compare
globular clusters, it is thus necessary to take into account the different stages of evolution of both the objects and
their environments of formation. Since the clusters form in a very evolving environment (evolution of the Universe and
the dynamical environment of the parent galaxy), the basic properties of different clusters are related to each other by
some evolutionary pattern. In particular, the dust and gas from which the stars of the globular clusters form have been
"polluted" (enriched in heavy elements) by more ancient stars, being field stars or belonging to other globular
clusters. This results in a kind of transmission with modification process, which justifies a priori the use of
cladistics. It must be clear that this is not a "descent with modification" in the sense that there is no replication.
But evolution does nevertheless create diversity. We are dealing with phylogeny (relationships between species), not
with genealogy (relationships between individuals). Since a multivariate classification of globular clusters is not yet
available, we assume in the present work that each cluster represents a “species” that will have to be defined later
on. 

As our work on galaxies has shown us, it is important to remove parameters that are redundant. Since previous studies of
the manifold of Galactic globular clusters have shown that 4 parameters are sufficient to describe their diversity, we
selected 4 parameters, 3 of which are intrinsic characteristics of the environment of formation. The fourth one, age, is
particular in the sense that it does not inform on the conditions when the clusters formed, and is not discriminant for
clustering because it evolves similarly for any cluster (parallel evolution). However, age is useful to rank the
clusters within each group. Consequently, we applied to age a weight half that of the other three parameters. In
addition, a stepmatrix was employed to impose the irreversibility constraint on the age parameter (age can only
increase). In contrast to multivariate distance methods, undocumented values are not a problem in cladistics
analyses. This is why the seven galaxies that have no age determination (see Fig.\ref{age_fe2fig} and
Table~\ref{tableprop}) have not been excluded in our work.

In this paper, we use parsimony as the optimisation criterion. This works as follows. One first builds a matrix with
values of the four parameters for all clusters. The values for each parameter are discretized into 10 bins representing
supposedly evolutionary states. Discretization of continuous variables is quite a complex problem, especially in
the evolutionary context (see e.g. \cite{GOL,TFB09}). The choice of the number of bins cannot be made in a
simple objective way. Here, we took equal-width bins, and considered a compromise between an adequate sampling
of continuous
variables and the uncertainties on the measurements. The first constraint is given by the software (32 in this case).
The second one would a priori give a lower limit of something like total range / uncertainty, but Shannon's theorem would
multiply this by 2. Hence, 10 bins would account for about 20\% measurement errors, which is rather large, but
\cite{RB06} do not provide precise error estimates, especially for logTe. Even so, border effects always imply that
some objects could belong to a bin or its neighbour, a process that add some more artificial noise. The best way to
avoid this effect is to make several analyses with different number of bins and check that the result does not depend on
this number. We have done this for 3, 5, 8, 10, 12, 15 and 20, and the result is identical, to at most one misplaced
cluster, for numbers higher than 8. For 3, no structure is found, and for 5 bins the groups are not well defined.

Then, all possible arrangements of clusters on a tree are constructed, and using the
discretized matrix, the total number of state changes is computed for each tree. The most parsimonious tree is finally
selected. If several such trees are found, then a consensus (strict or majority rule) tree is built. The whole procedure
is computerised since the number of arrangements is here very large. The result is a diversification scenario that
should be confronted to other knowledge and parameters. Maximum parsimony heuristic searches were performed using the
PAUP*4.0b10 (\cite{paup}) package. The results were interpreted with the help of the Mesquite software
(\cite{mesquite}). 

The tree presented in Fig. 1 is a majority rule consensus tree of 20000 trees, the strict consensus tree showing exactly the same three groups but with group 1 and 2 slightly less resolved. To further assess the robustness of the tree, it was not possible to make bootstrapping due to the irreversibility constraint on the age parameter, and it would not have been very significant with only 4 parameters. We performed other analyses using only 3 parameters, excluding the age. They all gave essentially the same three groups, but they were individually slightly less resolved, as expected. All these convergences yield strong confidence on the tree shown in Fig. 1. In the end, the most important point is the astrophysical interpretation we are able to give of the results. 

On Fig. 1, the tree is rooted with group 2 as an outgroup. This is not strictly necessary in cladistics, and we find here the same three groups whatever the root chosen or even on the unrooted tree. But we know that a low metallicity is an ancestral state for stars in general, this is why we have chosen group 2 a posteriori because it has a homogeneously low metallicity.

\label{lastpage}
\end{document}